\documentstyle[twocolumn,prb,aps,floats,epsf]{revtex} 

\begin{document}
\twocolumn[\hsize\textwidth\columnwidth\hsize\csname 
@twocolumnfalse\endcsname

\title{Zeeman effects on the impurity-induced
resonances in $d$-wave superconductors}
\author{Claudio Grimaldi} 
\address{\'Ecole Polytechnique F\'ed\'erale de Lausanne,
D\'epartement de Microtechnique IPM, CH-1015 Lausanne, Switzerland.}

\date{\today} 
\maketitle 

\begin{abstract}
It is shown how the resonant states induced by a single
spinless impurity in a $d_{x^2-y^2}$-wave superconductor evolve under the
effect of an applied Zeeman magnetic field. Moreover, it is demonstrated 
that the spin-orbit coupling to the impurity potential can have important and
characteristic effects on the resonant states and their response to the
Zeeman field, especially when the impurity is close to the unitary limit.
For zero or very small spin-orbit interaction, the resonant states
becomes Zeeman splitted by the magnetic field while when the spin-orbit
coupling is important, new low-lying resonances arise which do not
show any Zeeman splitting.
\\
PACS number(s): 74.20.-z, 71.70.Ej, 74.62.Dh
\end{abstract}

\vskip 2pc] 

\narrowtext

Recent scanning tunneling microscopy (STM) measurements in BSCCO high-$T_c$
superconductors have recorded clear images of quasiparticle resonant
states around intrinsic defects \cite{hudson} and individual Au \cite{yazda}
and Zn \cite{pan} impurity atoms.
A common result of these different measurements is that, in average,
the resonances are observed just below the Fermi energy, indicating a
resonant energy $\omega_0$ much smaller than the superconducting energy
gap $\Delta$ ($|\omega_0|\sim \Delta/30$). For the case of Zn-impurity
doping, the spatial dependence of the resonant states exhibits a large
signal at the impurity site with local maxima on the second neighbour Cu sites
({\it i. e.} along the node-gap directions) followed by somewhat weaker
peaks along the directions of the gap maxima.\cite{pan}

Low-lying quasiparticle resonant states have been predicted to occur in
$d_{x^2-y^2}$-wave superconductors doped with spinless impurities close
to the unitary limit.\cite{balatsky} According to such model, the resonance at
$\omega_0\simeq 1.5$ meV for Zn-substituted samples implies a scattering 
parameter $c=1/\pi N_F V_{\rm imp}$ of about $0.1$,\cite{pan}
where $V_{\rm imp}$ is the $s$-wave
impurity potential and $N_F$ is the normal-state density of states 
for each spin species at the Fermi level. 
Similar values of $c$ are estimated for Au impurities
and generic intrinsic defects.\cite{hudson,yazda} 
Furthermore, the observed power-law decay 
$\bar{G}(r)\sim 1/r^{1.97}$ of the angle-averaged differential conductance
$\bar{G}(r)$ at large distances $r$ from the impurity site \cite{hudson}
is in very good accord with $\bar{G}(r)\sim 1/r^2$, 
predicted in Ref.\onlinecite{balatsky}.

Despite of the agreements between theory and experiments, the spatial 
dependence of the resonant states reported in Ref.\onlinecite{pan} is quite ad odds
with that expected by strong spinless impurities. These, in fact, would 
generate, in addition to the peak at the impurity site, a fourfold symmetric
signal with maxima along the directions where the gap is 
fully opened.\cite{balatsky,haas}

The images recorded by the STM measurements appear therefore rotated by
$\pi /4$ with respect to those resulting from a spinless quasi-unitary
impurity potential. Recently, it has been proposed that a blocking effect
of the BiO and SrO layers between the tunneling tip and the CuO$_2$ layer
would actually give rise to the same spatial dependence recorder in
experiments.\cite{zhu} On the other hand,
according to a recent theoretical analysis,\cite{polko} the observed
spatial dependence would rather arise from a local antiferromagnetic
spin re-arrangement induced by a nominal zero-spin weak impurity. From
this perspective, the Zn-atom behaves effectively as a magnetic impurity
with non-local coupling to the charge carriers leading to the X-shaped
geometry of the resonant state. Such a picture would be consistent also
with the presence of unpaired $S=1/2$ moments in the vicinity of Zn atoms
as observed by NMR experiments.\cite{NMR}

From the above discussion and the contrasting claims reported in recent 
literature, it appears that the problem of deciding whether nominal spinless
impurities behaves as non-magnetic or effectively magnetic scattering centers
in high-$T_c$ cuprates is still an open issue.
However, models based on purely non-magnetic impurity potentials predict
resonant states quite sensitive to impurity strength and 
charge carrier doping,\cite{balatsky,flatte}
while the picture proposed in Ref.\onlinecite{polko} has been claimed to 
yield results much more robust. 
This qualitative difference could therefore be used as a
tool discriminating between the two pictures by, for example, studying the 
response to some external applied perturbation.

The aim of this paper is two-fold.
First, it is shown how the resonant states induced by a single
spinless impurity in a $d_{x^2-y^2}$-wave superconductor evolve under the
effect of an applied Zeeman magnetic field. Second, it is demonstrated 
that the spin-orbit coupling to the impurity potential becomes especially
important when the impurity is close to the unitary limit,\cite{grima1}
and can have important effects on the resonant states and their response
to a Zeeman field. 
According to whether the spin-orbit
scattering is irrelevant or not, the Zeeman response of the resonant
state behaves in two distinct ways. For zero or very small spin-orbit
interaction, the resonances become Zeeman splitted and the spatial dependence
can change from electron-like to hole-like for already quite small values
of the imposed magnetic field. On the contrary, when the spin-orbit coupling 
to the quasi-unitary impurity becomes relevant, new low-lying 
sharp resonances arise
which do not show Zeeman splitting. In this latter case, the spatial
dependence at fixed energy can be made to change from electron-like to
a novel, spin-orbit induced, symmetry for a suitable value of the Zeeman field.

The differential conductance recorded in a STM experiment is proportional
to the local density of states (LDOS) $N({\bf r},\omega)=-(1/\pi) {\rm Im}
[G_{11}^{\rm R}({\bf r},{\bf r};\omega)+
G_{22}^{\rm R}({\bf r},{\bf r};\omega)]$, 
where $G^{\rm R}({\bf r},{\bf r};\omega)$
is the retarded $4\times 4$ matrix Green's function defined in the 
particle-hole--spin space and ${\bf r}$ is the vector position with 
respect to the impurity located at the origin. The $(11)$ and $(22)$
components refer to the two pseudospin states of the quasiparticles.
For the single impurity case, $G^{\rm R}({\bf r},{\bf r};\omega)$ is
obtained by the Fourier transform of $G({\bf k},{\bf k}';\omega)=
\delta_{{\bf k},{\bf k}'}G_0({\bf k},\omega)+
G_0({\bf k},\omega)T({\bf k},{\bf k}';\omega)G_0({\bf k}',\omega)$,
where $G_0({\bf k},\omega)$ is the Green's 
function for the pure $d_{x^2-y^2}$-wave superconductor and 
$T({\bf k},{\bf k}';\omega)$ is the $T$-matrix associated to the impurity
scattering:
$T({\bf k},{\bf k}';\omega)=V({\bf k},{\bf k}')+
\sum_{{\bf k}''}V({\bf k},{\bf k}'')G_0({\bf k}'',\omega)
T({\bf k}'',{\bf k}';\omega)$, where 
$V({\bf k},{\bf k}')$ is the impurity potential in momentum space.
In the following, an external magnetic field ${\bf H}$ is assumed to be 
directed parallel to the conducting $x$-$y$ plane, for example 
${\bf H}\parallel \hat{{\bf x}}$, and the spins are quantized 
along the direction of ${\bf H}$.
Under the assumption of strong two-dimensionality, the coupling of the 
planar magnetic field to the quasiparticle spins becomes predominant over the
coupling to the quasiparticle orbital motion.\cite{makifulde}
In the limiting case for
which the orbital coupling can be neglected,\cite{samokhin} the 
Green's function $G_0({\bf k},\omega)$ is simply given by
$G_0^{-1}({\bf k},\omega)=\omega-\epsilon({\bf k})\rho_3-
\Delta({\bf k})\rho_2\tau_2+h\rho_3\tau_3$, where $\epsilon({\bf k})$ is
the quasiparticle dispersion, $\Delta({\bf k})\equiv 
\Delta(\phi)=\Delta \cos(2\phi)$ is the
gap function, and $h=\mu_{\rm B} H$ is the Zeeman energy ($\mu_{\rm B}\equiv$
Bohr magneton). The Pauli matrices $\rho_i$ and $\tau_j$ ($i,j=1,2,3$) 
act on the particle-hole and spin subspaces, respectively. 
For sufficiently low values of $h/\Delta$, the Fulde-Ferrel-Larkin-Ovchinnikov 
state can be ignored, and the effect of $H$ is merely to split 
the pseudo-spin degeneracy of the quasiparticle excitations.\cite{yang,won,note1}

The impurity atom is assumed to have a simple $\delta$-function potential:
$V_{\rm imp}({\bf r})=V_{\rm imp}\delta({\bf r})$. According to the
Elliott-Yafet theory,\cite{EY}
the spin-orbit coupling to $V_{\rm imp}$ can be modelled
as $V_{\rm so}({\bf r})=(\delta g/k_F^2) [\mbox{\boldmath$\nabla$} 
V_{\rm imp}({\bf r})\times{\bf p}]
\cdot\mbox{\boldmath$\sigma$}$, 
where ${\bf p}=-i\mbox{\boldmath$\nabla$}$ and $\mbox{\boldmath$\sigma$}$ 
are respectively the momentum and spin operators, $k_F$ is the Fermi momentum
and $\delta g$ is of the order of the shift of the 
$g$-factor.\cite{makifulde,EY}
Here, $\delta g$ is treated as a free parameter, however not exceeding
$\delta g\simeq 0.1-0.2$, which is the expected order of magnitude for
CuO$_2$ systems.\cite{note2} Since the charge carriers are confined to move on the
$x$-$y$ plane, only the $\sigma_z$ component of $V_{\rm so}({\bf r})$ is
nonzero. Hence, in the particle-hole spin subspace, the total impurity 
potential $V({\bf k},{\bf k}')$ reduces to $V({\bf k},{\bf k}')=
V_{\rm imp}\rho_3+i\,\delta g V_{\rm imp}
[\hat{{\bf k}}\times\hat{{\bf k}}']_z\tau_1$.

\begin{figure}
\epsfxsize=18pc
\epsfbox{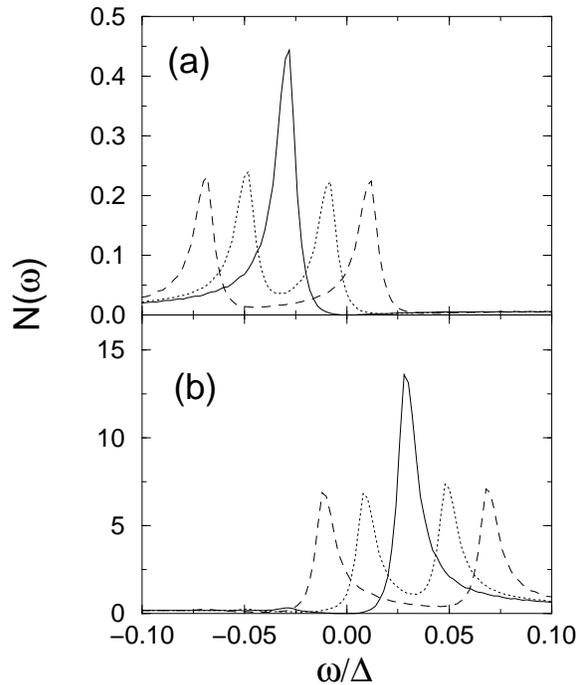}
\caption{LDOS without spin-orbit interaction as a function of
$\omega/\Delta$ for $c=0.08$ and $h=0$ (solid lines), $h/\Delta=0.02$
(dotted lines), and $h/\Delta=0.04$ (dashed lines).
(a): LDOS at the impurity site ${\bf r}=(0,0)$. (b): LDOS
along the direction of gap maxima ${\bf r}=(4/k_F,0)$}
\label{fig1}
\end{figure}

The ${\bf k}\times{\bf k}'$ dependence of the spin-orbit contribution permits
to decouple the $T$-matrix into two components: $T({\bf k},{\bf k}';\omega)=
T_{\rm imp}(\omega)+T_{\rm so}({\bf k},{\bf k}';\omega)$, where
$T_{\rm imp}(\omega)=V_{\rm imp}\rho_3+\sum_{{\bf k}}V_{\rm imp}\rho_3
G_0({\bf k},\omega)T_{\rm imp}(\omega)$ is the usual $T$-matrix for the scalar
potential and:
\begin{eqnarray}
&&T_{\rm so}({\bf k},{\bf k}';\omega)=
i\delta g V_{\rm imp}[\hat{{\bf k}}\times\hat{{\bf k}}']_z\tau_1 \nonumber \\
&&+i\delta g V_{\rm imp}\sum_{{\bf k}''}[\hat{{\bf k}}\times\hat{{\bf k}}'']_z
\tau_1G_0({\bf k}'',\omega)
T_{\rm so}({\bf k}'',{\bf k}';\omega), 
\label{tso}
\end{eqnarray}
is the $T$-matrix for the spin-orbit coupling to the impurity.\cite{grima1}
The resulting LDOS is therefore the sum of three contributions:
$N({\bf r},\omega)=N_0(\omega)+\delta N_{\rm imp}({\bf r},\omega)+
\delta N_{\rm so}({\bf r},\omega)$. 
From now on, the LDOS contributions are given in units of the normal-state
DOS summed over the two spin directions, $2N_F$, and particle-hole 
symmetry is assumed.
Hence, $N_0(\omega)=[{\rm Im}g_0(\omega_+)+
{\rm Im}g_0(\omega_-)]/2$, 
where $g_0(\omega_{\pm})=\int\frac{d\phi}{2\pi}\omega/
[\Delta(\phi)^2-\omega_{\pm}^2]^{1/2}$ and $\omega_{\pm}=\omega\mp h$,
is the LDOS for the pure superconductor, while 
$\delta N_{{\rm imp},({\rm so})}({\bf r},\omega)=
\delta N_{{\rm imp},({\rm so})}^+({\bf r},\omega)+
\delta N_{{\rm imp},({\rm so})}^-({\bf r},\omega)$ is the LDOS contribution
induced by the interaction with the impurity (spin-orbit) potential. 
The impurity part of the LDOS is just the superposition of
the zero-field LDOS reported in Ref.\onlinecite{balatsky} shifted by $\pm h$:
\begin{equation}
\label{Nimp}
\delta N_{\rm imp}^{\pm}({\bf r},\omega)=-\frac{1}{2}{\rm Im}
\left[\frac{g_0({\bf r},\omega_{\pm})^2}{g_0(\omega_{\pm})+c}+
\frac{f_0({\bf r},\omega_{\pm})^2}{g_0(\omega_{\pm})-c}\right],
\end{equation}
where
\begin{equation}
\label{gofo}
\left[\begin{array}{c} 
g_0({\bf r},\omega_{\pm}) \\
f_0({\bf r},\omega_{\pm}) \end{array}\right]
=\int\frac{d\phi}{2\pi}
\frac{e^{i{\bf k}_F\cdot{\bf r}}}
{[\Delta(\phi)^2-\omega_{\pm}^2]^{1/2}}
\left[\begin{array}{c} 
\omega_{\pm} \\
\Delta(\phi) \end{array}\right].
\end{equation}
A crucial effect of the Zeeman magnetic field is that the poles
arising from the denominators of Eq.(\ref{Nimp}) are splitted by $h$.
In fact, for small values of $c$ and $h$, the energy resonance $\omega_0(h)$
is simply $\omega_0(h)\simeq \omega_0 \pm h$, where 
$\omega_0\simeq\Delta (c\pi/2)/
\ln(8/\pi c)$ is the resonance energy for $h=0$.\cite{balatsky}
Hence, for quasi-unitary scattering ($\omega_0\ll \Delta$), already quite small
values of $h$ compared to $\Delta$ are sufficient to deeply modify
the impurity-induced resonance. 

This is clearly seen in Fig.\ref{fig1} where the $\omega$-dependence of $N_0(\omega)+
\delta N_{\rm imp}({\bf r},\omega)$ is plotted for $h/\Delta =0, 0.02, 0.04$.
The value of the scattering parameter, $c=0.08$, has been chosen in
order to reproduce a zero-field maximum signal on the impurity site, 
${\bf r}={\bf 0}$ (Fig.\ref{fig1}a), for $\omega/\Delta=-0.03$, {\it i. e.} 
the resonant energy reported in Ref.\cite{pan}. For nonzero values of $h$,
the resonance becomes Zeeman splitted in good agreement with 
$\omega_0(h)\simeq \omega_0 \pm h$. This is also true for the LDOS signals
away from ${\bf r}={\bf 0}$, as it is shown in Fig.\ref{fig1}b where the LDOS is
plotted for ${\bf r}=(4/k_F,0)$ ({\it i. e.}, along the direction
of the gap maxima). 

\begin{figure}
\epsfxsize=18pc
\epsfbox{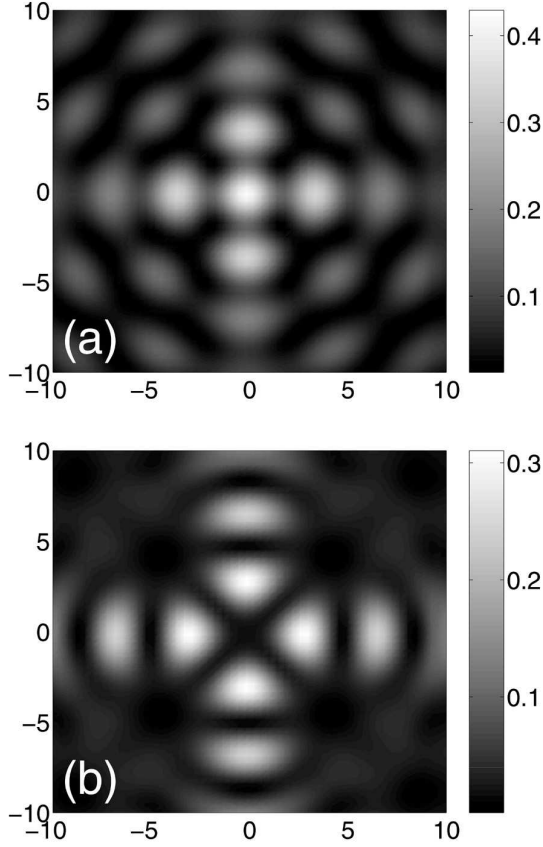}
\caption{Spatial dependence of the LDOS 
 without spin-orbit interaction
for $c=0.08$ and $\omega/\Delta=-0.03$. (a): $h=0$. (b): $h/\Delta=0.02$.}
\label{fig2}
\end{figure}

The spatial dependence of the LDOS as a function of $h$ for
$\omega/\Delta=-0.03$ is shown in Fig.\ref{fig2}.  The pattern shown in
Fig.\ref{fig2}a ($h=0$) closely resembles the spatial dependence obtained by
Haas and Maki,\cite{haas} and it is characteristic of an electron-like
bound state. However, already for $h/\Delta=0.02$ (Fig.\ref{fig2}b), which for 
$\Delta=44$ meV \cite{pan} corresponds to a magnetic field of about $15$ T, 
the resonance acquires a predominant hole-character, signalled by the
contemporary suppression of the central peak at ${\rm r}={\bf 0}$ and 
the signals along the gap-node directions. This pattern is equal to
that reported in Fig.\ref{fig2} of Ref.\onlinecite{haas}, but here it
has been obtained without
reversing the sign of $\omega$. As can be also inferred from Fig.\ref{fig1},
higher values of $h$ modify the intensity of the signal, 
but its spatial dependence remains equal to that of Fig.\ref{fig2}b.
It is important to stress that Fig.\ref{fig2} refers to the spatial
dependence on the CuO$_2$ layer and no blocking effect has been
considered. 

Let us consider now under which conditions the results of Figs 1 and 2 
are modified by the presence of spin-orbit coupling to the impurity.
The spin-orbit $T$-matrix contribution, Eq.(\ref{tso}), is obtained by
setting $T_{\rm so}({\bf k},{\bf k}';\omega)=i\delta g V_{\rm imp}
[\hat{{\bf k}}\times{\bf t}(\hat{{\bf k}}',\omega)]_z\tau_1$, where
${\bf t}(\hat{{\bf k}},\omega)=\hat{{\bf k}}+i\,\delta g V_{\rm imp}
\sum_{{\bf k}'}\hat{{\bf k}}'\tau_1G_0({\bf k}',\omega)
[\hat{{\bf k}}'\times{\bf t}(\hat{{\bf k}},\omega)]_z$.\cite{grima1} 
The equation for
${\bf t}(\hat{{\bf k}},\omega)$ is easily solved in terms of its components,
$t_x$ and $t_y$, and after some algebra the resulting spin-orbit 
part of the LDOS reduces to:
\begin{eqnarray}
\delta N_{\rm so}^{\pm}({\bf r},\omega)=&-&\frac{1}{2}{\rm Im}
\{A^{\pm}(\omega)[g_{\rm c}({\bf r},\omega_{\pm})^2+
g_{\rm s}({\bf r},\omega_{\pm})^2 \nonumber \\
&+&f_{\rm c}({\bf r},\omega_{\pm})^2+
f_{\rm s}({\bf r},\omega_{\pm})^2]\} \nonumber \\
&-&{\rm Im} \{B^{\pm}(\omega)
[g_{\rm s}({\bf r},\omega_{\pm})f_{\rm s}({\bf r},\omega_{\pm}) \nonumber \\
&-&g_{\rm c}({\bf r},\omega_{\pm})f_{\rm c}({\bf r},\omega_{\pm})]\},
\label{nso}
\end{eqnarray}
where
\begin{eqnarray}
\label{A}
A^{\pm}(\omega)&=&\frac{c_{\rm so}^2 g(\omega_{\mp})+[f(\omega_{\mp})^2-
g(\omega_{\mp})^2] g(\omega_{\pm})}{D(\omega)}, \\
\label{B}
B^{\pm}(\omega)&=&\frac{c_{\rm so}^2 f(\omega_{\mp})-[f(\omega_{\mp})^2-
g(\omega_{\mp})^2] f(\omega_{\pm})}{D(\omega)},
\end{eqnarray}
\begin{eqnarray}
\label{D}
D(\omega)=&&\{c_{\rm so}^2-[f(\omega_-)+g(\omega_-)]
[f(\omega_+)+g(\omega_+)]\} \nonumber \\
&&\times \{c_{\rm so}^2-[f(\omega_-)-g(\omega_-)]
[f(\omega_+)-g(\omega_+)]\}
\end{eqnarray}
where $c_{\rm so}=1/(\pi N_F\delta g V_{\rm imp})=c/\delta g$ and:
\begin{equation}
\label{gf}
\left[\begin{array}{c} 
g(\omega_{\pm}) \\
f(\omega_{\pm}) \end{array}\right]
=\int\frac{d\phi}{2\pi}
\frac{\sin(\phi)^2}
{[\Delta(\phi)^2-\omega_{\pm}^2]^{1/2}}
\left[\begin{array}{c} 
\omega_{\pm} \\
\Delta(\phi) \end{array}\right],
\end{equation}
\begin{equation}
\label{gcfc}
\left[\begin{array}{c} 
g_{\rm c}({\bf r},\omega_{\pm}) \\
f_{\rm c}({\bf r},\omega_{\pm}) \end{array}\right]
=\int\frac{d\phi}{2\pi}
\frac{\cos(\phi)\,e^{i{\bf k}_F\cdot{\bf r}}}
{[\Delta(\phi)^2-\omega_{\pm}^2]^{1/2}}
\left[\begin{array}{c} 
\omega_{\pm} \\
\Delta(\phi) \end{array}\right].
\end{equation}
The expressions for $g_{\rm s}({\bf r},\omega_{\pm})$ and 
$f_{\rm s}({\bf r},\omega_{\pm})$ are obtained
by replacing $\cos(\phi)$ with $\sin(\phi)$ in the right-hand side
of Eq.(\ref{gcfc}).

There are two important general features characteristic of the spin-orbit LDOS.
First, as it can be inferred from Eqs.(\ref{nso}-\ref{gcfc}),
the spin-orbit LDOS vanishes at the impurity site:
$\delta N_{\rm so}^{\pm}({\bf 0},\omega)=0$. Moreover, at $h=0$,
$\delta N_{\rm so}({\bf r},0)=0$ for every ${\bf r}$. This is due to the
${\bf k}\times {\bf k}'$ factor appearing in Eq.(\ref{tso}) which makes
the spin-orbit coupling particle-hole symmetry conserving. 
The second important feature is that the spin-orbit interaction can induce
additional resonances driven by the zeroes of Eq.(\ref{D}). 
Without entering too much into details,
the main feature of the spin-orbit poles is that, at $h=0$,
$D(\omega)$ vanishes at $\omega=0$ when
$c_{\rm so}= 1/\pi$.\cite{note3} 
Away from this limit the poles acquire
a finite imaginary part and move rapidly towards high energies. 
Note however that since $\delta N_{\rm so}({\bf r},0)=0$, the
resonance becomes sharper as $c_{\rm so}\rightarrow 1/\pi$
without reducing to a $\delta$-function at $\omega=0$ for $c_{\rm so}= 1/\pi$.
Finally, due to the spin-mixing processes of the spin-orbit interaction, 
the effect of $h$ is not merely a Zeeman split of the zero-field poles.

\begin{figure}
\epsfxsize=18pc
\epsfbox{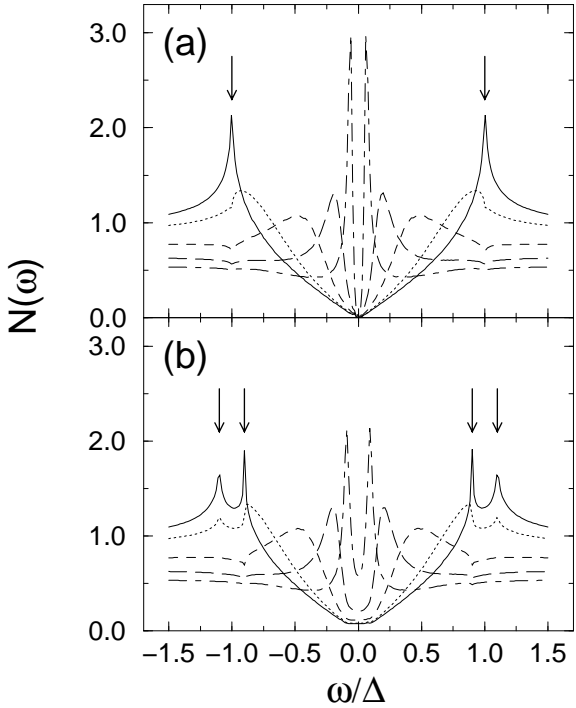}
\caption{Effect of the spin-orbit contribution to the LDOS
at ${\bf r}=(2/k_F,2/k_F)$
for $c=0.08$ and different values of the spin-orbit parameter $\delta g$
for $h=0$ (a) and $h/\Delta=0.1$ (b). Solid lines: $\delta g=0.01$;
dotted lines: $\delta g=0.05$;  dashed lines: $\delta g=0.1$;
long dashed lines: $\delta g=0.15$; dot-dashed lines: $\delta g=0.2$.
The arrows indicate the positions of the coherence peaks for
$\delta g=0.01$.}
\label{fig3}
\end{figure}

These features are demonstrated in Fig.\ref{fig3} where the total LDOS
including the spin-orbit contribution is plotted as a function of $\omega$
for ${\bf r}=(2/k_F,2/k_F)$, $c=0.08$ and different values of $\delta g$.
For $h=0$ (Fig.\ref{fig3}a) and $\delta g=0.01$ the presence of the
coherence peaks at $\omega=\pm \Delta$ indicate that the LDOS is very
close to that of a $d$-wave superconductor without impurities. However,
as $\delta g$ is enhanced, the coherence peaks are depleted and
a symmetric broad resonance develops and moves towards
low energies with contemporary reduction of its peak-width. 
The symmetry with respect to $\omega=0$ merely reflects the particle-hole
symmetry conservation of the spin-orbit interaction. At 
$\delta g=0.2$, the coherence peaks at $\omega=\pm \Delta$ are
completely suppressed and a sharp resonance is built at 
$\omega=\pm \omega_{\rm so}$ with $\omega_{\rm so}\ll \Delta$.
The origin of such low-lying resonances stem from the poles
of Eq.(\ref{D}). Note in fact that for $\delta g=0.2$ the value 
of the spin-orbit scattering parameter, $c_{\rm so}=c/\delta g=0.4$, 
nearly fulfils
the condition $c_{\rm so}=1/\pi$ for which, as discussed above,
the spin-orbit $T$-matrix has a pole at $\omega_{\rm so}=0$ with vanishing 
imaginary part. The effect of
the magnetic field is shown in Fig.\ref{fig3}b, where the LDOS is plotted for
$h/\Delta=0.1$ and for the same set of $\delta g$ values of
Fig.\ref{fig3}a. As expected, for $\delta g=0.01$ the coherence peaks
at the gap edge are splitted by $\pm h$. However, for higher values 
of $\delta g$, the low-lying spin-orbit resonances do not show Zeeman
splitting because of the presence of important spin-flip processes.

\begin{figure}
\epsfxsize=18pc
\epsfbox{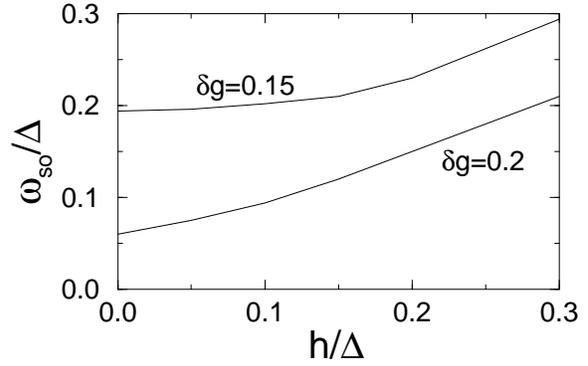}
\caption{Spin-orbit resonant energies $\omega_{\rm so}$ for
$c=0.08$ as a function of the external field.}
\label{fig4}
\end{figure}

Although the low-lying spin-orbit resonances are not Zeeman
splitted, they nevertheless show some dependence on $h$. This is
shown in Fig.\ref{fig4} where $\omega_{\rm so}$ is plotted as a
function of $h$ for $\delta g=0.15$ and $\delta g=0.2$. Note however
that the $h$-dependence of $\omega_{\rm so}$ is rather weak, at least
for low values of $h$.

\begin{figure}
\epsfxsize=18pc
\epsfbox{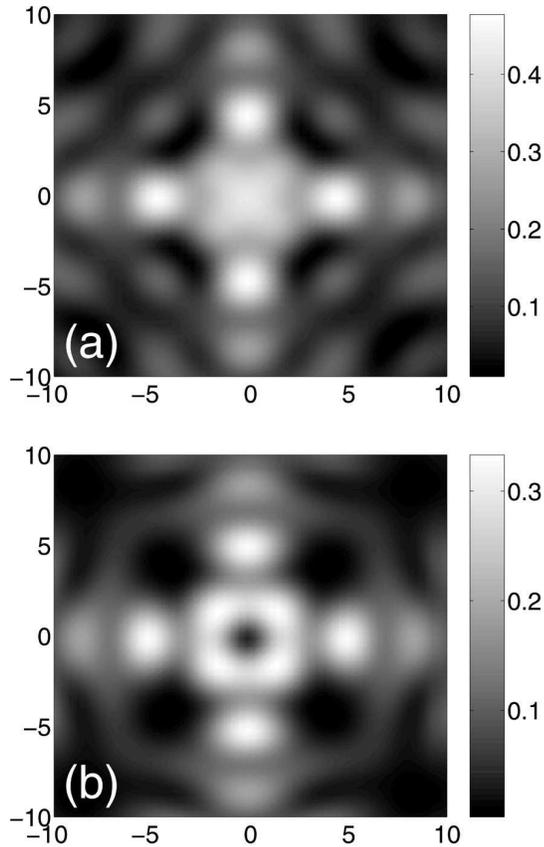}
\caption{Spatial dependence of the LDOS with spin-orbit interaction
for $c=0.08$, $\delta g=0.2$ and $\omega/\Delta=-0.03$. 
(a): $h=0$. (b): $h/\Delta=0.02$. }
\label{fig5}
\end{figure}

The spatial dependence of the total LDOS with $\delta g=0.2$
is plotted in Fig.\ref{fig5} for the same parameters of Fig.\ref{fig2}
($c=0.08$ and $\omega/\Delta =-0.03$). For $h=0$ (Fig.\ref{fig5}a) the poles
of the spin-orbit $T$-matrix are at energies higher than $\omega=-0.03$
and the spatial dependence of the LDOS resembles closely that of Fig.\ref{fig2}a
where $\delta g=0$. Note however that the weight of the central peak
is somewhat extended along the diagonals leading to an X-shaped geometry.
For $h/\Delta=0.02$ (Fig.\ref{fig5}b) the spatial dependence is radically different
from that shown in Fig.\ref{fig2}b. Now, a signal arises along the diagonals
in the vicinity of ${\bf r}={\bf 0}$ with a contemporary shift of the
peaks in the $(\pm 1,0)$ and $(0,\pm1)$ directions at higher distances from
the impurity site. This particular geometry of the LDOS is characteristic
of the spin-orbit coupling to the impurity and, as inferred from Fig.\ref{fig3}, it
can be obtained also at $h=0$ when $\omega$ is close to the spin-orbit
$T$-matrix poles.

In summary, two possible scenarios can be drawn about the Zeeman field effects
on the LDOS of a $d_{x^2-y^2}$-wave superconductor 
around a quasi-unitary impurity atom. First, if the spin-orbit coupling is
absent or weak ($c_{\rm so}\gg 1/\pi$), the imposed magnetic field splits
the quasiparticle resonance peaks by $\pm h$ and, at fixed energy $\omega$,
the spatial dependence can be modified from electron-like to hole-like.
Second, if the spin-orbit scattering is sufficiently strong 
($c_{\rm so}\simeq 1/\pi$), the LDOS acquires a novel off-site and 
particle-hole symmetric resonance at low energies which do not show
Zeeman splitting at $h\neq 0$. Which of these two possibilities is actually
realized would bring important informations on the nature of the
resonant states in high-$T_c$ superconductors.

A last remark regards the possibility of having very high (small) values of
$\delta g$ ($c$) is such a way that $c_{\rm so}\ll 1$. In fact, when
$c_{\rm so}\rightarrow 0$, equations (\ref{A}) and (\ref{B})
reduce to $A^{\pm}(\omega)=g(\omega_{\pm})/[f(\omega_{\pm})^2-
g(\omega_{\pm})^2]$ and
$B^{\pm}(\omega)=-f(\omega_{\pm})/[f(\omega_{\pm})^2-
g(\omega_{\pm})^2]$, respectively. In this case, 
$\delta N_{\rm so}^{\pm}(\omega)$ does no longer contains spin-mixed
terms and the two spin channels are perfectly decoupled. 
As explained in Ref.\onlinecite{grima2}, this situation is due to
the fact that, as long as the charge carriers are confined to move in the
$x-y$ plane, the spin-orbit impurity operator
commutes with $\sigma_z$. For very strong spin-orbit interaction, it is
found that also in the presence of an external magnetic field perpendicular
to the $z$ direction, the (singlet) Cooper pairs are formed by electrons with
opposite spins in the $z$-direction and the spin-rigidity of the 
superconducting condensate is efficient against spin-flip transitions
induced by the magnetic field. 

{\it Note added} - As shown in a recent publication [H. Tsuchiura, Y. Tanaka,
M. Ogata, and S. Kashiwaya, Phys. Rev. Lett. {\bf 84}, 3165 (2000)]
a Zeeman splitting of the quasiparticle resonances can be
induced also by classical magnetic impurities.


\begin{references}
\bibitem{note1}
Note that, at zero temperature, the gap suppression due to the Zeeman
field is $\ln (\Delta/\Delta_0)=-(h/\Delta)^3/3$, where $\Delta_0$ is the
order parameter for $h=0$ (see Refs\cite{yang,won}). Here, $h$ is varied 
up to $h=0.3\Delta \ll \Delta$, so that no distinction is made
between $\Delta$ and $\Delta_0$.
\bibitem{note3}
At $h=0$, Eq.(\ref{D}) reduces to $D(\omega)=
[c_{\rm so}^2-(f(\omega)+g(\omega))^2][c_{\rm so}^2-(f(\omega)-g(\omega))^2]$.
Since $g(\omega=0)=0$ and $f(\omega=0)=-1/\pi$, $D(\omega=0)=0$ if
$c_{\rm so}=\pm 1/\pi$.
\end{references}
\end{document}